\documentclass{mem}
\usepackage{natbib}\usepackage{txfonts}
\usepackage{graphicx}
\usepackage[a4paper]{hyperref}
\idline{75}{1}
\begin{document}
\def\teff{\mbox{$T\rm_{eff }$}}
\def\kms{\mbox{$\mathrm {km s}^{-1}$}}
\def\logg{\mbox{$\log g$}}
\def\mturb{\mbox{$\xi\rm_{turb}$}}
\def\fconv{\mbox{$f\rm_{conv}$}}

\title{Different convection models in ATLAS
}

   \subtitle{}

\author{Barry Smalley}

\institute{Astrophysics Group,
Keele University,
Staffordshire,
ST5 5BG,
United Kingdom \newline
\email{bs@astro.keele.ac.uk}
}

\abstract{Convection is an important phenomenon in the atmospheres of A-type and
cooler stars. A description of convection in ATLAS models is presented, together
with details of how it is specified in model calculations. The effects of changing
the treatment of convection on model structures and how this affects observable
quantities are discussed. The role of microturbulence is examined, and its link
to velocity fields within the atmosphere. Far from being free parameters,
mixing-length and microturbulence should be constrained in model calculations.


}
\maketitle{}

\section{Introduction}

The gross properties of a star, such as broad-band colours and flux
distributions, are significantly affected by the effects of convection in
stars. Consequently, our modelling of convection in stellar atmosphere models
can significantly alter our interpretation of observed phenomena.

\subsection{Mixing-Length Theory}

\begin{figure}[t!]
\resizebox{\hsize}{!}{\includegraphics[clip=true]{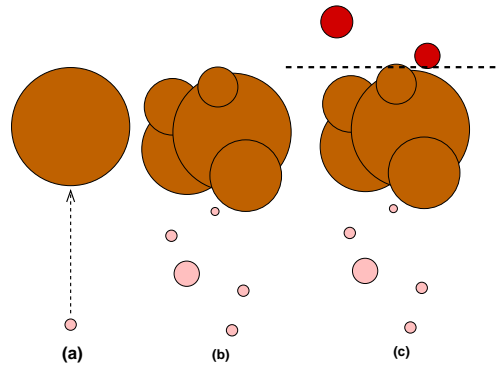}}
\caption{\footnotesize Schematic bubble representations of convection
treatments. In mixing-length theory (a), a single bubble rises within the
atmosphere, while in turbulent convection bubbles of varying sizes rise (b).
In (c) we have overshooting above the convection zone.}
\label{bubble}
\end{figure}

Convection in stellar atmospheres is usually based on mixing-length theory
(MLT) of \citet{BV58}. In this model a single bubble of gas rises a certain
mixing-length ($l/H$) before dispersing (Fig.~\ref{bubble}a). The problems with
this theory is that it is clearly too simple and that the mixing-length is a
totally free parameter. In their discussion of the ATLAS6 models \citet{RK78}
found discrepancies between theoretical and observed $uvby$ colours which might
be the result of inappropriate treatment of convection within the models.
Subsequently, several attempts have been undertaken to improve the situation.
\citet{LLK82}, for example, introduced ``horizontally averaged opacity'' and a
``variable mixing length'' which improved the match with observed $uvby$
colours, but did not remove all the discrepancies.

\subsection{The Canuto \& Mazzitelli (CM) Model}

\cite{CM91,CM92} proposed a turbulent model of convection in order to overcome
one of the most basic short-comings of MLT, namely that a single convective
element (or ``bubble'' or ``eddy'') responsible for the transport of all the
energy due to convection. This new model accounts for eddies of various sizes
that interact with each other (Fig.~\ref{bubble}b). The CM convection model was
implemented in the ATLAS9 code by \cite{KUP96} and has no user adjustable free
parameters. An improved variant is the self-consistent (CGM) method of
\cite{CGM96}.

\subsection{Convective Overshooting}

Convective bubbles rise above the convections zone into the stable regions
(Fig.~\ref{bubble}c). This is called overshooting, and should be present in our
model atmosphere calculations. The ATLAS9 models introduced an ``approximate
overshooting'' which has not been without its critics (see \citealt{CGK97} for
full details). The following quote from Kurucz' web site is aptly summarizes
the situation:

\begin{quote} {\it convective models use an overshooting approximation that moves
flux higher in the atmosphere above the top of the nominal convection zone.
Many people do not like this approximation and want a pure unphysical
mixing-length convection instead of an impure unphysical mixing-length
convection} \end{quote}

\subsection{Atmospheric Structure}

\begin{figure}[t!]
\resizebox{\hsize}{!}{\includegraphics[clip=true]{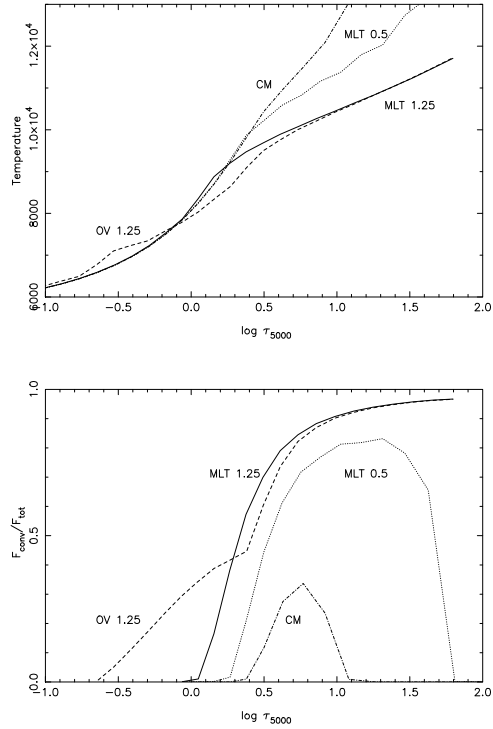}}
	\caption{\footnotesize Models with {\teff} =
	7000~K and {\logg} = 4.0 showing differences arising from changing the
	treatment of convection. The upper panel shows the variation of
	temperature with optical depth, while the lower panel shows the
	convective flux as function of optical depth}
\label{structure}
\end{figure}

At {\teff} = 8000K, CM gives essentially radiative temperature gradient with
significantly less convective flux than MLT, while  approximate overshooting
introduces flux into higher layers \citep{HEI+02}. Figure~\ref{structure}
shows the situation for slightly cooler models (\teff=7000~K). The CM model
remains close to the radiative temperature gradient. MLT gives more convective
flux than CM, even when $l/H$ = 0.5. Overshooting produces an excess of
convective flux in higher layers, which produces a noticeable bump in the
temperature-depth relation compared to MLT without overshooting.

\section{The ATLAS CONVECTION Card}

A single control card specifies how the treatment of convection within ATLAS.

If the card is not present in the ATLAS input, convection is turned off. A
couple of variables have default values which should be noted:
mixing-length {\tt MIXLTH} = 1 and amount of overshooting {\tt OVERWT} = 1.

\begin{description}
\item{\tt \bf CONVECTION OFF}

Ensures that convection is turned off for the model calculation. Note that this
sets {\tt MIXLTH} to 1, but leaves {\tt OVERWT} unchanged.

\item{\tt \bf CONVECTION ON $<$MIXLTH$>$}

Turns on convection within the model calculation. The value of
mixing-length {\tt MIXLTH} is set to the user specified value,
but {\tt OVERWT} is not modified. Hence {\tt OVERWT} will be the default
value of 1 and {\em approximate overshooting is enabled by default}.

\item{\tt \bf CONVECTION OVER $<$MIXLTH$>$ $<$OVERWT$>$}

This variant of the control card, which allows full control over approximate
convective overshooting. See \cite{CGK97} for details of convective
overshooting.

Note that setting {\tt OVERWT} = 0 turns off approximate overshooting.

\item{\tt \bf CONVECTION OVER $<$MIXLTH$>$ $<$OVERWT$>$ $<$NCONV$>$}

ATLAS puts the constraint that the convective flux (\fconv) must be zero above
layer {\tt NRHOX/2} (i.e. higher than the middle layer in the model atmosphere
desk) This was originally introduced to remove numerical artifacts. The value
is usually a good number, except for coolest models, where constraint generates
a jump in T($\tau_{\rm Ross}$) for {\teff} $\le$ 4000K \citep{CAS05}. In order
to alleviate this problem, {\tt NCONV} introduced into a version of ATLAS by
Castelli to allow user to specify the layer above which {\fconv} is surely
zero.  The default value is 36. This is {\tt NRHOX}/2 for {\tt NRHOX} = 72
which is typical number of layers used in a model calculation.

\end{description}

\subsection{The CM and CGM routines}

Drop-in replacements for the {\tt CONVEC} and {\tt TCORR} subroutines
were implemented by \cite{KUP96}.

Usage is similar to standard ATLAS version: {\tt \bf CONVECTION ON
$<$MIXLTH$>$}. However, there are differences depending on whether the CM or
CGM routines are used:

\begin{description}

\item[CM] In this case {\tt MIXLTH} not actually used, but {\em must} be set to
$>$ 0 in order for the routine to work correctly. This is to ensure that rest
of the ATLAS codes knows that convection is running.

\item[CGM] Here {\tt MIXLTH} has the meaning of $\alpha$* (see \citealt{CGM96}
for details). The standard value used is 0.09.

\end{description}

\section{Testing convection models}

None of the current 1d models of convection are totally satisfactory. 2d and 3d
numerical simulations are producing impressive results \citep{FS04}, as are
improved analytical 1d treatments \citep{KUP04}. However, in order to be
confident that the current generation of ATLAS models are producing reliable
results, we need to know how good the treatment of convection is in ATLAS and
the limitations.

\subsection{uvby photometry}

\cite{SK97} compared the predicted $uvby$ colours for the CM model with that
from the standard ATLAS9 MLT models with and without ``approximate
overshooting''. Comparison against fundamental {\teff} and {\logg} stars
revealed that the CM models gave better agreement than MLT without
overshooting. Models with overshooting were clearly discrepant. This result was
further supported by stars with \teff\ obtained from the Infrared Flux Method
(IRFM) and {\logg} from stellar evolutionary models. However, some
discrepancies still remained, including a ``bump'' around 6500~K in the {\logg}
obtained for the Hyades and continued problems with the Str\"{o}mgen $m_0$
index. Similar conclusions were found by \cite{SCH99} using Geneva photometry.

\subsection{The $m_0$ index}

\begin{figure}[t!]
\resizebox{\hsize}{!}{\includegraphics[clip=true]{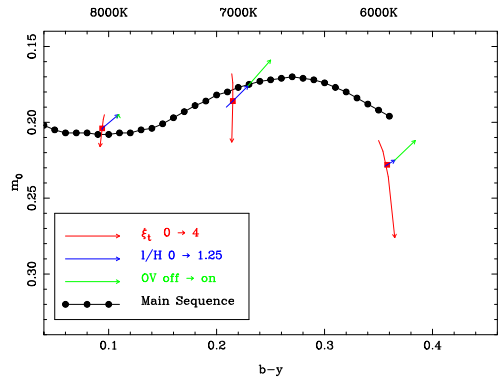}}

 \caption{\footnotesize The variation of $m_0$ index with $b-y$ showing the sensitivity to
 microturbulence (\mturb), mixing-length ($l/H$) and ``approximate
 overshooting''. At each temperature the model with $\log g$ = 4.0, \mturb\ =
 2 \kms\ and $l/H$ = 0.5 is denoted by a square. The arrows indicate the effect
 of varying \mturb, $l/H$ and including overshooting (for $l/H$ = 1.25). The
 \cite{PE80} main-sequence is included for reference.}

\label{m0}
\end{figure}

The $m_0$ index is sensitive to metallicity and microturbulence, but also
convection efficiency \citep{RK78,SK97}. Inefficient convection (CM and MLT
$l/H$ $\sim$ 0.5) clearly works in the domain of the A stars down to $b-y$
$\sim$ 0.20 ($\sim$7000~K, F0). For cooler stars, convection becomes more
efficient and substantive within the atmosphere and higher values of
mixing-length and lower microturbulent velocities would be required to fit the
observed $m_0$ indices of the main sequence (Fig.~\ref{m0}). The 2d numerical
radiation hydrodynamics calculations of \cite{LFS99} indicate a rise in
mixing-length from $l/H$ $\sim$1.3 at 7000~K to $l/H$ = 1.6 for the Sun
(5777~K), while a much lower $l/H$ $\sim$ 0.5 was found at 8000~K
\citep{FRE95}. This is in agreement with that implied by $m_0$ index. However,
around 6000~K there still remains a significant discrepancy, which could only
be reduced by invoking the approximate overshooting option (Fig.~\ref{m0}).
None of the convection models used in classical model atmospheres allows for
the reproduction of the $m_0$ index, unless $l/H$ and the amount of
``approximate overshooting'' are varied over the H-R Diagram.

\subsection{Stellar Fluxes}

The stellar flux distribution is influenced by the effects of convection on the
atmospheric structure of the star. As we have seen with photometric colours,
these effects have a clearly observable signature. Hence, high-precision
stellar flux measurements should provide significant and useful information on
convection.

\cite{LLK82} presented a study of convective model stellar atmospheres using a
modified mixing-length theory. They found small, systematic differences in the
optical fluxes. Their figures also demonstrate that convection can have a
measurable effect on stellar fluxes.

\begin{figure}[t!]
\resizebox{\hsize}{!}{\includegraphics[clip=true]{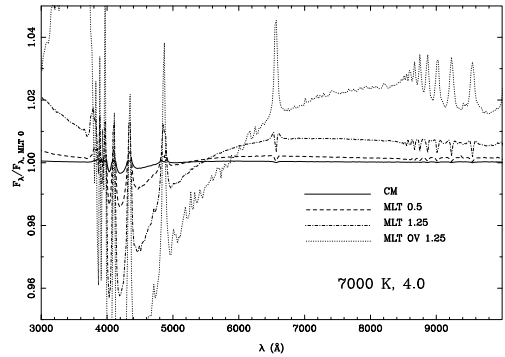}}
	\caption{\footnotesize Fluxes for CM models and MLT models, with $l/H$ = 0.5 and 1.25,
 compared to that for a model with zero convection. Their are noticeable
 differences, especially in the region 4000 $\sim$ 5000\AA, and the effect of
 overshooting is considerable.}
\label{fluxes}
\end{figure}

Figure~\ref{fluxes} shows the effect of changing mixing length from 0, through
0.5 to 1.25 on the emergent flux for solar-composition models with \teff
=7000~K, \logg = 4.0 and \mturb = 2 \kms.  The differences are noticeable, with
the effect of overshooting being considerable.

\subsection{Balmer line profiles}

\begin{figure}[t!]
\resizebox{\hsize}{!}{\includegraphics[clip=true]{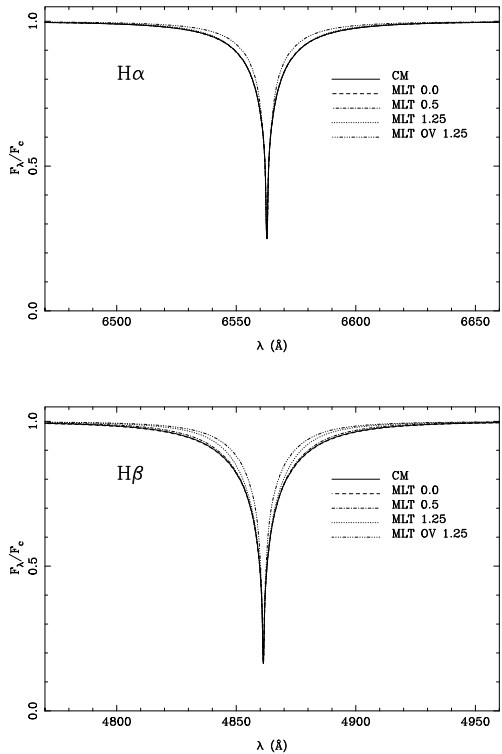}}
	\caption{\footnotesize The effects of convection on the predicted shape of Balmer
 profiles for models with \teff\ = 7000, $\log g$ = 4.0, [M/H] = 0.0 and
 \mturb\ = 2 \kms. $H\alpha$ (upper panel) is unaffected by the values of $l/H$, but
 sensitive to ``approximate overshooting'', while $H\beta$ (lower panel) is sensitive to
 both.}
\label{balmer}
\end{figure}

The temperature sensitivity of Balmer lines makes them an excellent diagnostic
tool for late A-type stars  and cooler \citep{GAR00}. However, as emphasised by
\cite{vM96}, the $H\alpha$ and $H\beta$ profiles behave differently due to
convection: $H\alpha$ is significantly less sensitive to mixing-length than
$H\beta$. Both profiles are, nevertheless, affected by the presence of
overshooting, with $H\beta$ being more influenced than $H\alpha$
(Fig.~\ref{balmer}). Since $H\alpha$ is formed higher in the atmosphere than
$H\beta$, Balmer lines profiles are a very good depth probe of stellar
atmospheres. Naturally, Balmer profiles are also affected by microturbulence,
metallicity and, for the hotter stars, surface gravity \citep{HEI+02}.

In their studies of H$\alpha$ and H$\beta$ profiles of A and F stars \cite{GKS99} and
\cite{SMA+02} found good agreement with fundamental stars for CM and MLT ($l/H$ 
$\sim$ 0.5) without approximate overshooting. However, \cite{GKS99} found $l/H$
= 1.25 gave better results for cooler stars ({\teff} $>$ 7000~K).

\section{Microturbulence}

Microturbulence is a free parameter introduced to allow abundances from weak
and strong lines to agree. It is an extra source of broadening, which is added
to thermal broadening of stellar lines. Physically, it is postulated as
small-scale turbulent motions within the atmosphere, where the size of the
turbulent elements is less than the unit optical depth \citep{GRA92}.

\begin{figure}[t!]
\resizebox{\hsize}{!}{\includegraphics[clip=true]{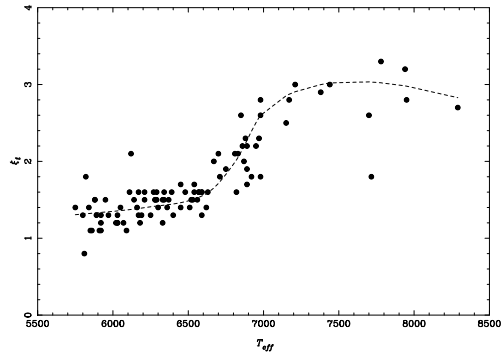}}
	\caption{\footnotesize The variation of microturbulence with effective temperature. Based on
 results of \cite{GGH01} for stars near the main sequence. The dashed
 line indicates the approximate variation with {\teff}. Note the
 apparent relatively abrupt change in behaviour between 6500 and 7000~K}
\label{micturb}
\end{figure}

Microturbulence does appear to vary with effective temperature. Several studies
have found that microturbulence appears to vary systematically with {\teff}
\citep{CHA70,NIS81,CB92,GGH01}. Fig.~\ref{micturb} shows the variation of
\mturb\ with \teff\ for near main-sequence stars ($\log g$ $>$ 4.0) based on
the results given by \cite{GGH01}. Microturbulence increases with increasing
{\teff}, peaking around mid-A type, before falling away to zero for B-type
stars. There is a relatively abrupt change in behaviour between 6500 and
7000~K, which is related to the change from weak subsurface convection to the
fully convective atmospheres of cooler stars.

\begin{figure*}[t!]
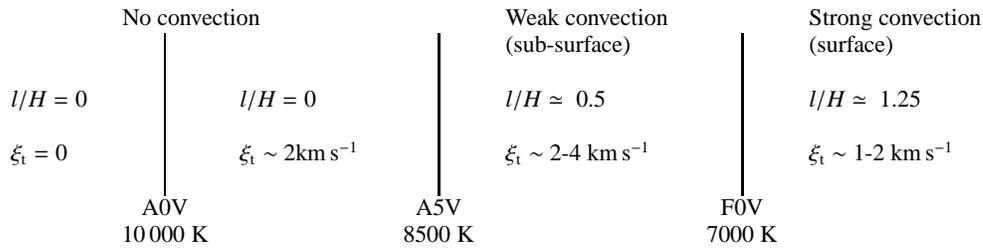

\begin{tabular*}{\textwidth}{p{1.05cm}cp{1.75cm}cp{2.25cm}cp{2.45cm}}
\multicolumn{3}{c}{No convection} && Weak convection && Strong convection \\
 &\vline&&\vline& (sub-surface) &\vline& (surface) \\
 &\vline&&\vline&&\vline& \\
$l/H = 0$ &\vline& $l/H = 0$ &\vline& $l/H \simeq\ 0.5$
&\vline& $l/H \simeq\ 1.25$ \\
 &\vline&&\vline&&\vline& \\
$\xi_{\rm t} = 0$ &\vline& $\xi_{\rm t} \sim$ 2km\,s$^{-1}$
&\vline& $\xi_{\rm t} \sim$ 2-4 km\,s$^{-1}$
&\vline& $\xi_{\rm t} \sim$ 1-2 km\,s$^{-1}$ \\
 &\vline&&\vline&&\vline& \\
&A0V&&A5V&&F0V& \\
&10\,000~K&&8500~K&&7000~K \\
\end{tabular*}
 \caption{The Convection Recipe for stars near the main sequence \citep{SMA04}}
 \label{recipe}
\end{figure*}

\subsection{Line Asymmetries}

Velocity fields are present in stellar atmospheres which can be measured using
line bisectors \citep{DRA87,GRA92}. Compared to solar-type stars, the line
bisectors in A-type stars are reversed, indicating small rising columns of hot
gas and larger cooler downdrafts \citep{LAN98}.  It is these motions that are
thought to be responsible, at least in part, for the existence of
microturbulence. In fact, 3d numerical simulations of solar granulation can
account for observed line profiles without the need for any microturbulence
\citep{ASP+00}. Similar results have been found for Procyon
\citep{GRA85,AP+02}, which is also a star with well-known physical parameters
(e.g. \citealt{KER+04}).

Numerical simulations avoid the need for a microturbulence free parameter
\citep{ASP+00}. The microturbulence of 1d is not turbulent motions,  but rather
velocity gradients within the atmosphere. Hence, microturbulence should no
longer a free parameter, but ought to be constrained within ATLAS model
calculations. Infact, Kurucz presented an empirical method for constraining
depth-dependent microturbuluce within ATLAS at this Workshop \citep{KUR05}.

\section{A Convection Recipe}

\cite{SMA04} presented a schematic variation of microturbulence and mixing
length with {\teff} for stars near the main sequence (Fig.~\ref{recipe}).
For stars hotter than A0 there is no convection or
significant microturbulence. For the early A-type stars there is essentially no
convection within the atmosphere, since the temperature gradient is radiative,
but there are velocity fields as indicated by the modest microturbulence
values. Velocity fields increase as we go through mid to late A-type stars, and
inefficient convection is required within the atmosphere. Once convection
becomes efficient (F-type and later) the value of microturbulence is found to
drop, while the mixing-length increases.

\section{Conclusions}

The effects of convection on the stellar atmospheric structure can be
successfully probed using a variety of observational diagnostics. The
combination of photometric colours and Balmer-line profiles has given us a
valuable insight into the nature of convection in stars. High quality
observations that are currently available and those that will be in the near
future, will enable further refinements in our theoretical models of convection
and turbulence in stellar atmospheres.

Overshooting is still an issue to be resolved, since there are clearly velocity
fields above the convection zone. While the ``approximate overshooting'' of
Kurucz appears to have been discounted by observational tests, there is clearly
the need for some sort of overshooting to be incorporated within model
atmosphere calculations.

\begin{acknowledgements}
The author would like to thank Fiorella Castelli and
Friedrich Kupka for useful discussions on convection and its implementation within
ATLAS.
\end{acknowledgements}

\bibliographystyle{aa}

\end{document}